# Protective Performance of Water-Based Epoxy Coating Containing ScCO$_2$ Synthesized Self-Doped Nanopolyaniline

M.R. Bagherzadeh[1][*], T. Mousavinejad[2][*], E. Akbarinezhad[1], A. Ghanbarzadeh[1]

1- Coating Research Center, Research Institute of Petroleum Industry (RIPI), Tehran 1485733111, Iran

Telephone: +98 (21) 48255212, Fax: +98 (21) 48255231, Email: bagherzadehmr@ripi.ir

2- Department of Chemistry, University of Puerto Rico, San Juan, PR 00931-3346, USA

Telephone: +1-787-3631401, Email: tahereh.mousavinejad@upr.edu

**Abstract**

In this research, corrosion properties of water-based epoxy coating is improved by adding self-doped nano polyaniline (SPAni) synthesized using supercritical carbon dioxide (ScCO$_2$) medium. The modified ScCO$_2$-synthesized self-doped nano polyaniline (SPAni) using water-based polyamidoamine hardener results formation of the water-based SPAni composite (Sc-WB). To obtain the water-based polyaniline epoxy coating (SP-WBE), the Sc-WB was mixed with epoxy resin in stoichiometric ratio. The formation of oxide layers and adhesion properties of SP-WBE at corrosive medium was evaluated using scanning electron microscopy (SEM) studies. The transmission electron microscopy (TEM) was used to observe the distribution and particles size of nanopolyaniline in the final dried film. The anti-corrosion performance of water-based epoxy coated sample (blank coated sample) and SP-WBE coated sample was studied using salt spray standard test according to ASTM B-117, electrochemical impedance spectroscopy (EIS) and adhesion tests. The results indicate protective performance of water-based epoxy coating is improved by adding self-doped polyaniline synthesized at ScCO$_2$ situation. The electrical resistance of the coatings containing nanopolyaniline and the blank sample was 1.8778E+7 $\Omega.cm^2$ and 2.512E+5 $\Omega.cm^2$ after 1800 hours of immersion in 3.5 Wt% NaCl aqua corrosive solutions.

# 1. Introduction

Conducting polymers (CPs) are widely used in various areas of research including electronics, solar cells, sensors, batteries and other energy storage devices, electromagnetic shielding devices and coatings [1-6]. The utilization of CPs as protective coatings against corrosion has been studied by many investigators because of restrictions on the use of coatings containing heavy metals, due to their serious environmental problems. Recently, most studies show that the use of industrial organic coatings based on CPs can provide meaningful protection efficiency for mild steel [7-9].

Among the available CPs, polyaniline (PAni) is found to be the most promising, because of its good environmental stability, low cost, excellent electrical properties, plastic nature, low cost and ease of synthesis comparing to other CPs [10-11].

One of the best explored fields of PAni applications is the corrosion protection of metals [12]. Popovic *et al*. [13] and Talo *et al*. [14] have shown that PAni/epoxy coatings exhibit much better protection behavior than single epoxy coatings. However, the industrial application of PAni due to its poor conductivity and low solubility in most common solvents has been limited. Chen *et al*. [15] reported the synthesis of water-soluble self-doped PAni. In order to improve its solubility in water and prevent dopant migration and at the same time to increase its stability and conductivity, PAni has been modified in the self-doped form by the introducing protic acids into the side chains.

Variety methods for the synthesis of PAni have been described, including solution polymerization, interfacial polymerization, seeding polymerization, electrochemical synthesis, template synthesis and polymerization in supercritical (Sc) [16]. Pham *et al*. [17]

reported the synthesis of PAni nanofibers in compressed $CO_2$ without use of a template or surfactant. It is demonstrated both high yield and small diameter of the PAni nanofibers could be obtained at the same time and the PAni nanofibers had one-dimensional crystalline structure without a hollow core, which gives high thermal stabilities.

Among various approaches used to prepare nanoPAni, $ScCO_2$ method uses a nontoxic, low cost, naturally abundant, nonflammable solvent with easily accessible supercritical conditions where $T_c = 31.1\ °C$ and $P_c = 73.8$ bar. Furthermore, physicochemical properties such as density and hence the solvent quality of $ScCO_2$ can be achieved with small changes in pressure and/or temperature [18-19].

Du *et al*. [20] developed a new route to prepare PAni microtubes via $ScCO_2$/aqueous interfacial polymerization. The synthesis is based on the chemical oxidative polymerization of aniline in an acidic environment, with ammonium peroxydisulfate (APS) as an oxidant and sodium dodecyl sulfate (SDS) surfactant as a template. It was demonstrated that well-dispersed PAni microtubes can be easily obtained and the diameter of PAni microtubes dictates their superior performance as chemical sensors.

In this study, the self-doped nano polyaniline (SPAni) was synthesized in $ScCO_2$ medium and used as a corrosion inhibitor additive to prepare a new water-based epoxy coating. To the best of our knowledge there are no previous reports in case of using synthesized SPAni in $ScCO_2$ medium into the water based hardener to increase corrosion property of epoxy coatings. The effect of the SPAni on adhesion and anticorrosion properties of water-based epoxy coating was investigated using standard methods such as salt spray, adhesion and electrochemical impedance spectroscopy (EIS).

## 2. Experimental

### 2.1. Materials

All reagents including aniline, 3-amino benzene sulfonic acid and ammonium persulfate were used as analytical grade and purchased from Merck Chemicals without any purification. To make the coatings, the water-based polyamidoamine hardener (RIPI-W.B.H.) [21] and the SPAni made in research institute of petroleum industries (RIPI) were used. Also the diglycidyl ether of bisphenol A epoxy resin (Epon 1001) purchased from Shell Company was used as binder.

### 2.2. Synthesis of self-doped nano polyaniline (SPAni)

The Aniline was distilled under reduced pressure and used as monomer. Copolymerization of aniline and 3-amino benzene sulfonic acid (SAN) was performed to obtain the SPAni. The ammonium persulfate (APS) was used as oxidizing agent. The aniline (0.0132 mol) and 3-amino benzene sulfonic acid (0.0044 mol) were added to 150 mL aqueous solution and the resulting dispersion was transferred into the high pressure reactor and heated up to 35°C (*see supporting information Fig. S1*). The achieved solution was stirred constantly for 15 min at 50 rpm. The APS agent (0.022 mol) was dissolved in 150 mL of distilled water and rapidly transferred into the reactor. The $CO_2$ gas was compressed into the reactor up to 100 bars and the reaction was carried out at 100 bars and 35 °C for 24 hr. The formed SPAni particles at the bottom of reactor under high pressure were added into the 250 mL of distilled water and filtered. The powdery substance was washed using deionized water for three times then dried at 50 °C under vacuum for 24 hr. The SPAni specifications are listed in Table I.

### 2.3. Preparation of water-based SPAni composite (Sc-WB)

To prepare the water-based SPAni composite, 0.3 gr SPAni, synthesized using *in-situ* polymerization in $ScCO_2$ medium, was diluted with water and sonicated for 2 minutes by an ultrasonic homogenizer (Hielscher UIP1000hd). Then, the suspension product was homogenized by adding 125 gr water-based hardener matrix (using a Polytron 6100 Kinematica homogenizer) for 15 min and the final sonication was carried out at ambient temperature for 5 min.

## 2.4. Characterization of Sc-WB

The structural characterization of SPAni, water-based hardener and Sc-WB was carried out using a Bruker IFS-88 FTIR spectrometer. The FTIR spectra were recorded in KBr pellets at room temperature in the range of 500 - 4000 $cm^{-1}$ with 4 $cm^{-1}$ resolution.

The particle size and distribution analysis for SPAni in water and RIPI-W.B.H. were studied by dynamic light scattering (DLS) using Pultron Nano Zetasizer instrument.

## 2.5. Preparation of water-based nanopolyaniline epoxy coating (SP-WBE)

The prepared Sc-WB composite (achieved at section 2.3) was added to the 337.5 gr epoxy resin and mixed together using mechanical mixer to make the SP-WBE. To study the effect of Sc-WB composite on final properties of coatings, a blank sample without nanoparticles was prepared. Afterwards, these mixtures were applied using air spray on near-white metal blasted steel plates (SSPC-SP10) to the average thickness of 80±5μm according to ASTM D7055. The curing process was done at room temperature for approximately one week.

## 2.6. Characterization of SP-WBE

Transmission electron microscopy (TEM) using Philips CM200 was carried out to distinguish dispersion of SPAni in water-based nanopolyaniline epoxy coatings.

The morphology of SP-WBE coating and coating-metal interface on cross section view after exposing the coated plates to the salt spray test medium for 1800 hours was done using HITACHI S4160 scanning electron microscope (SEM). For this purpose the coated plates after 1800 hours being in salt spray chamber was cut, mounted and mechanically polished using metallographic process.

The adhesion property of the coatings was measured according to ASTM D3359 before and after salt spray test while the salt spray test carried out according to ASTM B117 and interpreted using ASTM D1654 standard.

For further studies, the corrosion protection properties of coatings were evaluated by using electrochemical impedance spectroscopy (EIS) technique. EIS measurements were performed on specified area of coated mild steel panels using a Potentiostat/Galvanostat Autolab model 302N over a frequency range of 100 KHz to 10 mHz with ±20 mV amplitude of sinusoidal voltage at open circuit potential. For this studies, the three-electrode measurement was employed where the coated sample was working electrode, a platinum grid was auxiliary (counter) electrode and a saturated calomel electrode (SCE) was as reference one. Interpretation and simulation of impedance results were performed using the ZSim3.22 software.

## 3. Results and Discussion

### 3.1. Fourier transform Infrared (FT-IR) spectroscopy

Fig. 1 compares the typical FT-IR spectra of SPAni (1a), water-based hardener (1b) and Sc-WB composite (1c). The spectrum of the SPAni (Fig. 1a) shows bands at 1581 $cm^{-1}$ and 1493 $cm^{-1}$ which is the characteristic of the aromatic C=C double bond stretching of Quinoid and Benzenoid rings, respectively. The almost equal intensity for these two peaks indicates that the PAni could be in the emeraldine state [22]. The peaks located at 1035 $cm^{-1}$, 707 $cm^{-1}$ and 618 $cm^{-1}$ are due to O=S=O, S–O and C–S stretching in bipolaron structure, respectively. The peak at 1153 $cm^{-1}$ is due to C–H bonding vibration of protonated doped PAni and the peak at 1310 $cm^{-1}$ is related to C–N stretching in the bipolaron structure [23]. The SPAni and water-based hardener spectra at 3440 $cm^{-1}$ and 3403 $cm^{-1}$, respectively, show the wide peaks that indicate the presence of strongly primary alcohols and amines dipole groups such as –O–H and –N–H [22]. In case of Sc-WB, the corresponding peak (~3425 $cm^{-1}$) appears at similar energy but it is broader due to the presence of methylene groups in Sc-WB. In Fig. 1c, the characteristic peaks at 1462 cm$^{-1}$ and 1557 cm$^{-1}$ are slightly shifted to higher wave numbers in comparison with the water-based hardener. Therefore, from FTIR analysis it can be concluded that there is a strong interaction between SPAni and water-based hardener and most probably the water-based hardener particles are fully covered with SPAni.

### 3.2. Particle size study

The particle size distribution curves were determined by dynamic light scattering (DLS) for SPAni in water and RIPI-W.B.H (water-based hardener). The results are shown in Fig. 2 and Fig. 3.

Figure 2 illustrates the particle size distribution of the SPAni in water measured by DLS examination. The mean particle size of the SPAni dispersed in water is 2875 nm. It can be seen the size of particles in water is larger than their primary size before dissolving. To reduce the particle size by removing agglomerated particles, ultrasonic dispersion is adopted to ensure the SPAni can be well dispersed in the polymeric matrix. The dispersion size of the SPAni particles in water-based hardener is shown in Fig. 3 and it has a range of 2-3 nm. It indicates SPAni nanoparticles had a good interaction with water-based hardener and the molecules of hardener can encircle polyaniline particles to make a homogenized mixture with SPAni nanoparticles which is in good agreement with previous studies [24].

### 3.3. Scanning electron microscopy (SEM) analysis

The microstructure of synthesized SPAni was evaluated by SEM analysis and is illustrated in Fig. 4a. The micrograph demonstrates the flake morphology. SEM studies were also used to investigate the coatings adhesion and interaction of iron surface and coating at the interface and also to observe the iron oxide layer on the metal surface at the cross section (Fig 4. a and b.) The SEM analysis of samples was done after 1800 hours of exposure in a salt spray chamber according to ASTM B117. The SEM observation reveals the presence of oxide layer forming at the interface of metal-coating for both blank and SP-WBE coated samples. The thickness of oxide layer for the blank sample is 0.46-6.15µm while the oxide layer at the interface of SP-WBE coating forms a more uniform layer with 1.18-1.65 µm thickness.

**3.4. Transmission electron microscopy (TEM) analysis**

To determine distribution and size of SPAni flakes and particles in film the TEM examination was used. To prepare the TEM sample couple of drops of SPAni nanoparticles and water-based hardener and epoxy resin diluted mixture was dispensed on ultra-thin holy carbon copper grid. The TEM image is illustrated in Fig. 5. Based on the TEM studies of SPAni nanoparticles the average diameter of nanoparticles is 60 nm and a there is bridge-linked structure with a good distribution in the dry film.

**3.5. Salt spray and adhesion tests**

The iron plates coated by water-based epoxy after 1800 hours exposure time in salt spray test medium for blank and SP-WBE coated samples are shown in Fig. 6.

Visual investigation of corrosion confirms the existence of the areas with blisters and corrosion products on top of coating film for the blank sample. The rusted points and disbanding film can be observed on the surface of blank coated sample as shown in Fig. 6a. The White spots which are clearly observable on the surface of the sample indicates the diffusion of water, oxygen and corrosive ions such as chlorine through the coating to reach to the metal-coating interface to form corrosion products.

In case of the coating made by SPAni there is no blister or rust on the coating surface after 1800 hrs in salt spray test as it can be seen in Fig. 6b. In this sample several black spots are observed on the surface specially around the scratched area. However, after removing the coating film, there was no evidence of rust or other failures on the metal surface.

Investigating the coating appearance according to ASTM D1654 shows the presence of No.6 (medium) blisters and rusted points around the scratched area for the blank coated sample.

On the other hand for SP-WBE coated sample there is no blister or rust even on the scratched zone. Water penetration zone close to the scratched area was about 4-5 cm for the blank coated sample while it is about 5 mm for the SP-WBE coated sample.

The adhesion properties of coatings were investigated using Adhesion test according to ASTM D3359 method. Table II shows the results of adhesion test before and after 1800 hours exposure time to salt spray atmosphere. According to these results, the adhesion properties of blank coated sample decrease from 5B to 4B; in contrast, SP-WBE does not show considerable changes in adhesion after 1800 hours being in salt spray test. These results show that the adhesion and corrosion properties of water-based epoxy coatings are obviously improved by adding synthesized SPAni particles under $ScCO_2$ condition.

### 3.6. Electrochemical impedance spectroscopy (EIS)

The Bode plots for the steel plates coated with blank and SP-WBE coatings are shown in Fig. 7 and Fig. 8 at different immersion times in 3.5 Wt.% sodium chloride solution. Three-electrode measurement was used to carry out the EIS studies while the saturated calomel electrode (SCE) was selected as a reference electrode, the platinum grid as an auxiliary (counter) electrode (CE) and the coated steel specimen as a working electrode (WE) with a surface area of 4 $cm^2$. The equivalent circuits for EIS results that plotted in Fig. 7 and Fig. 8 are illustrated in Fig. 9.

The EIS studies for the specimens at the initial of immersion in 3.5 Wt.% NaCl solution indicateds extremely high resistance ($R_c$) for both SP-WBE and blank coated samples. The resistance ($R_c$) for the blank and SP-WBE coated samples as measured using EIS data fitting, calculated as 2.008E+8 $\Omega.cm^2$ and 6.435E+10 $\Omega.cm^2$, respectively. These results indicate that both coatings have a perfect resistance to water diffusion through the films structure at the beginning of immersion in corrosive solution and have very low probable paths for water molecules and other destructive ions to diffuse from coating-water interface to the metal-coating interface.

The EIS results after 1000 hours immersing in corrosive solution were fitted more to the two-capacitance loops where one appears at high frequencies and another at low frequencies. The first capacitance loop is due to the coating characteristics while the second one is related to the processes occurring at the coating-metal interface. This type of behavior is usually can be simulated by a porous film model equivalent circuit shown in Fig. 9.b. In this equivalent circuit, $R_s$ is the solution resistance, $C_c$ is the capacitance behavior of the coating, $R_c$ is the coating resistance, $C_{dl}$ is the capacitance behavior of the interfacial double layer and $R_{ct}$ is the charge transfer resistance. The simulated parameters which obtained by ZSim3.22d software is listed in Table III.

The impedance for the blank and SP-WBE coated samples after 1000 hours immersion in 3.5 Wt.% NaCl solution drops to 3.368E+5 $\Omega.cm^2$ and 1.425E+7 $\Omega.cm^2$, respectively as shown in Fig. 8 and Fig. 9. The EIS studies shows that the coating resistance ($R_c$) value for SP-WBE coated samples declines from 6.435E+10 $\Omega.cm^2$ to 1.425E+7 $\Omega.cm^2$ and then gradually increases to 1.8778E+7 $\Omega.cm^2$ while the sample remains in the solution from zero to 1000 hours and finally 1800 hours, respectively. This shows improving the corrosion protection properties for SP-WBE coated samples. The capacitance value ($C_c$) for blank coated sample decreases from 2.338E-8 $F.cm^{-2}$ to 3.389E-8 $F.cm^{-2}$ while it drops from 3.355E-10 $F.cm^{-2}$ to 6.283E-10 $F.cm^{-2}$ for SP-WBE sample when the immersion time in 3.5 Wt.% NaCl solution increases from zero to 1000 hours. This indicates that the protective property increases by increasing immersion time to 1000 hours.

In fact, due to diffusion of water molecules and destructive ions and atoms such as chloride through the coating, the coating resistance decreases while the coating capacitance increases with increasing immersion time.

The charge transfer resistance ($R_{ct}$) value for blank and SP-WBE coated samples were measured as 1.134E+6 $\Omega.cm^2$ and 1.576E+9 $\Omega.cm^2$, respectively. It indicates that the water molecules diffuse into the coatings and subsequently the corrosion phenomena can occur at the metal-coating interface after 1000 hours immersing in 3.5 Wt.% NaCl water solution.

By increasing immersion time from 1000 hours to 1800 hours, SP-WBE coated sample shows increasing in coating resistance ($R_c$) from 1.425E+7 $\Omega.cm^2$ to 1.8778E+7 $\Omega.cm^2$. Increasing in immersion time increases the corrosion product volumes at the metal-coating interface and subsequently the probability for blocking the diffusion paths through the coating increases. This phenomenon can boost the coating resistance after increasing immersion time from 1000 Hours to 1800 hours.

The coating capacitance ($C_c$) value for blank and SP-WBE coated samples were measured as 5.217E-8 $F.cm^{-2}$ and 6.342E-10 $F.cm^{-2}$ after 1800 hours immersion in 3.5 Wt.% NaCl solution, respectively. It can be noticed that the variation in coating capacitance ($C_c$) values for SP-WBE coated sample is lower than blank coated one after immersion. The charge transfer resistance ($R_{ct}$) value for blank and SP-WBE samples decreases to 4.527E+5 $\Omega.cm^2$ and 1.558E+9 $\Omega.cm^2$, respectively while the immersion time increases from 1000 hours to 1800 hours. The corrosion products include iron ions such as $Fe^{2+}$ or $Fe^{3+}$ especially at the metal-coating interface. By increasing iron ions the cathodic reaction can occur as $Fe^{3+} + e^- = Fe^{2+}$ therefore for continuing dissolving of iron atoms from the metal surface there is no need for oxygen atoms to transfer through the diffusion paths in the films to feed the cathodic reaction. As the result, after increasing immersion time from 1000 hours to 1800 hours, $R_{ct}$ decreases while the $R_c$ increases.

## 4. Conclusions

1- Low cost self-doped nano polyaniline (SPAni) has successfully been synthesized in a $ScCO_2$ medium.

2- To the best of our knowledge as the first time we report preparation of an anti-corrosion water-based epoxy coating using 0.07% of $ScCO_2$ synthesized SPAni under $ScCO_2$ condition which was added to the water-based hardener (RIPI-W.B.H.) matrix using mechanical mixing. The SPAni particles size in the water-based hardener varies from 2 to 3 nm and fully separated into the film.

3- The FT-IR measurement indicates that the SPAni has strong interactions with the water-based hardener.

4- According to the SEM studies the synthesized SPAni has a flake shape morphology that causes increasing the barrier properties by increasing the length of diffusion paths for corrosive ions and oxygen through the coating. One of the most significant results is to reach stable oxide layer, which was formed in coating-metal interface by adding the SPAni which can protect the steel surface against corrosion phenomenon.

5- Salt spray test, adhesion studies and EIS measurements show that adding 0.07% of synthesized SPAni in $ScCO_2$ medium in water-based epoxy coating significantly improves corrosion resistance.

Table I. Self-doped nano polyaniline (SPAni) specification

| Properties | Results |
|---|---|
| Appearance | Flake |
| Color | Dark green |
| Conductivity (S/Cm) | 0.051 |
| Particle Size (μm) | 1-2 |
| Purity (%) | >99 |

Table II. The results of adhesion test according to ASTM D3359

| Coating | Result | |
|---|---|---|
| | Before salt spray | After salt spray |
| Blank | 5B | 4B |
| SP-WBE | 5B | 5B |

Table III. Impedance parameters of blank and SP-WBE coated steel in 3.5% NaCl

| Time (hours) | Ref | | | | SP-WBE | | | | |
|---|---|---|---|---|---|---|---|---|---|
| | $R_c$ | $C_c$ | $R_{ct}$ | $C_{dl}$ | $R_c$ | $C_c$ | $R_{ct}$ | CPE | |
| | | | | | | | | Y(S.sec^n) | n |
| 0 | 2.008E+8 | 2.338E-8 | - | - | 6.435E+10 | 3.355E-10 | - | - | - |
| 1000 | 3.368E+5 | 3.389E-8 | 1.134E+6 | 9.795E-8 | 1.425E+7 | 6.283E-10 | 1.576E+9 | 4.310E-10 | 0.96044 |
| 1800 | 2.512E+5 | 5.217E-8 | 4.527E+5 | 1.828E-7 | 1.8778E+7 | 6.342E-10 | 1.558E+9 | 4.138E-10 | 0.98614 |

**Figures caption**

Figure 1. FTIR spectrum of a) SPAni, b) water-based hardener and c) SP-WBE

Figure 2. Particle size distribution of the SPAni in water

Figure 3. Particle size distribution of the SPAni in water-based hardener matrix

Figure 4. SEM pictures of a) SPAni, b) blank coating and c) SP-WBE coating

Figure 5. TEM of SPAni particle size in dry film coating

Figure 6. Image of plates coated with a) blank coating and b) SP-WBE coating after 1800 h exposure to salt spray medium

Figure 7. Bode plots of blank coated sample at different immersion time in 3.5% NaCl

Figure 8. Bode plots of SP-WBE at different immersion time in 3.5% NaCl

Figure 9. Equivalent circuit for blank and SP-WBE coated samples at a) t=0 and b) t=1000 and 1800 hours exposure in 3.5% NaCl

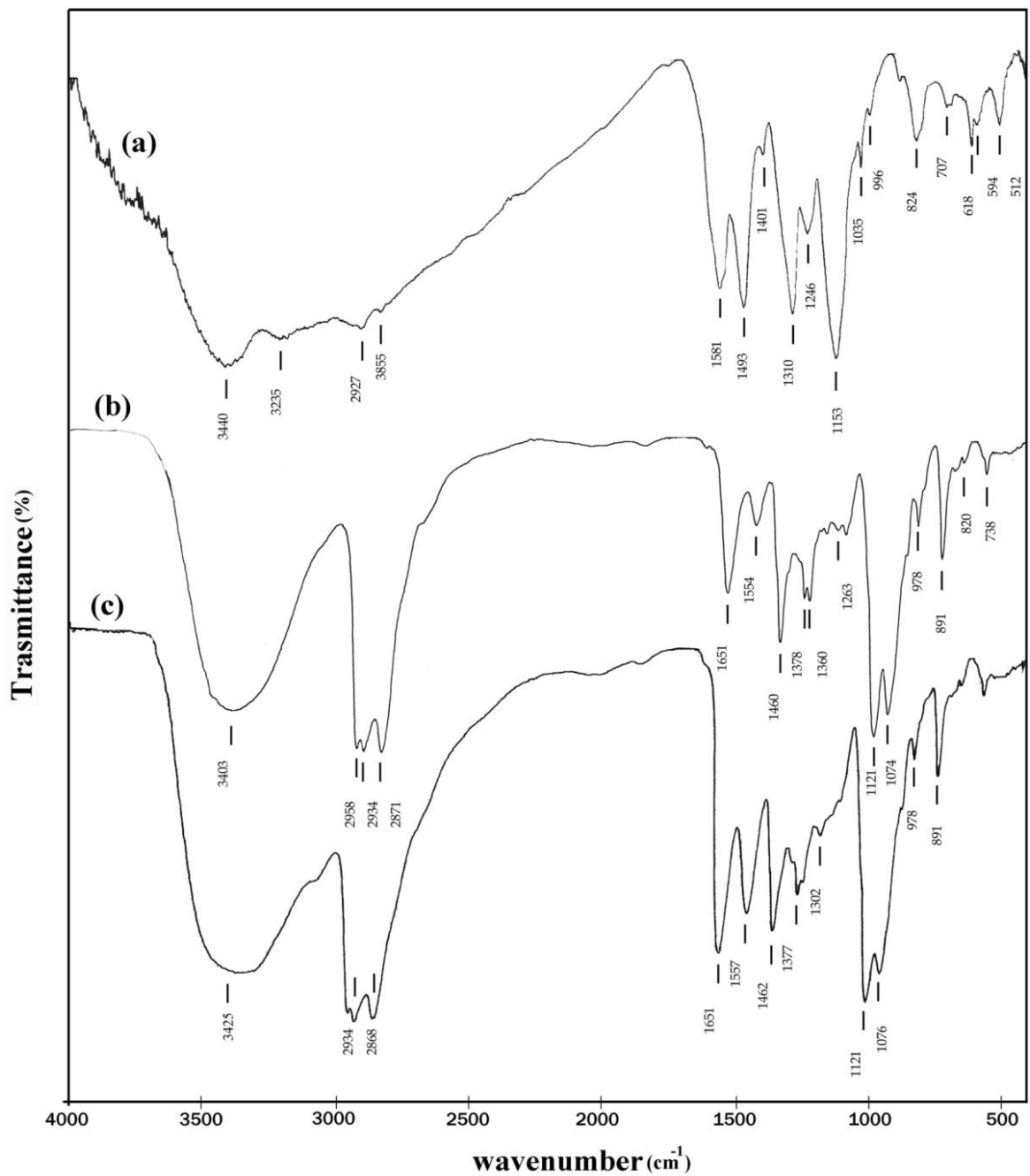

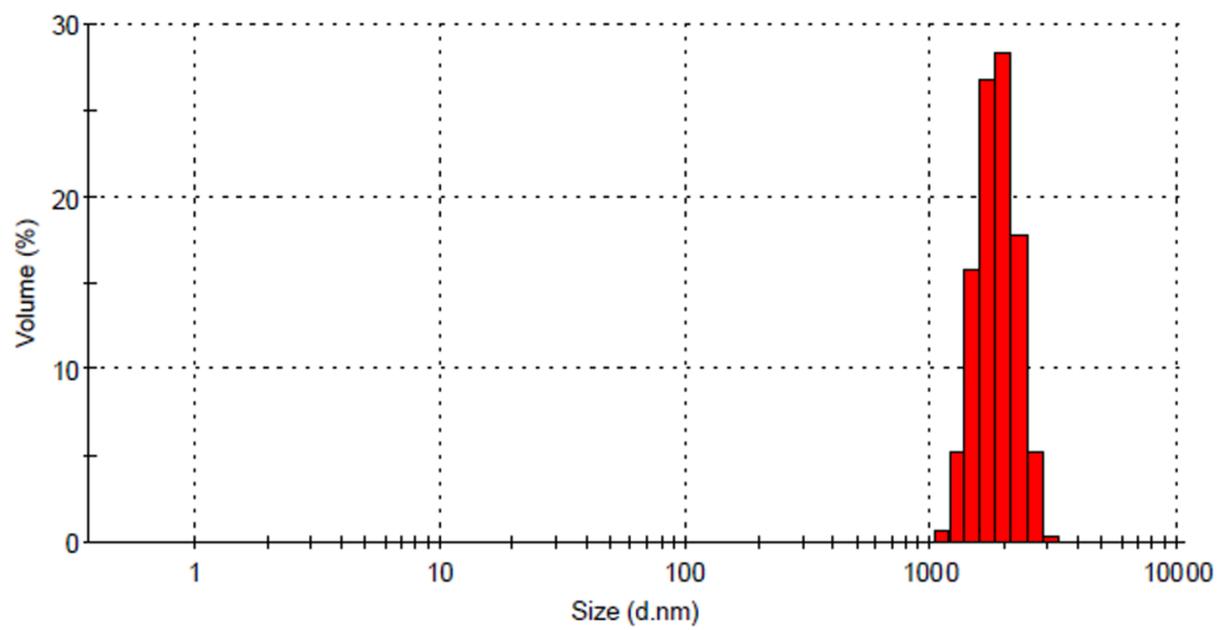

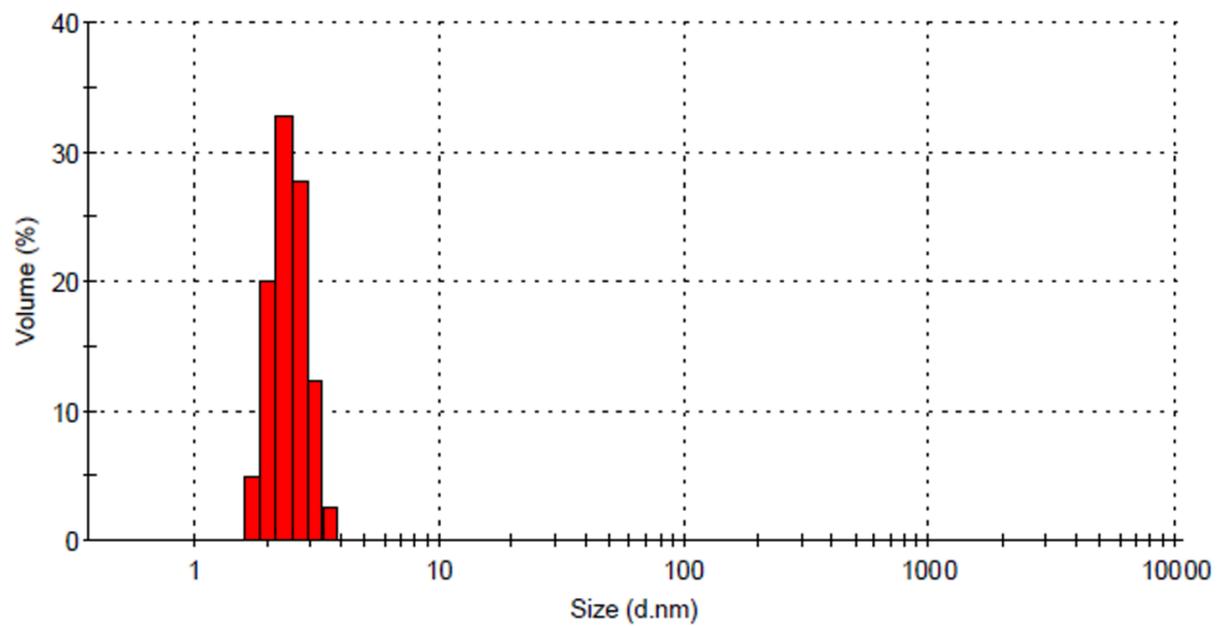

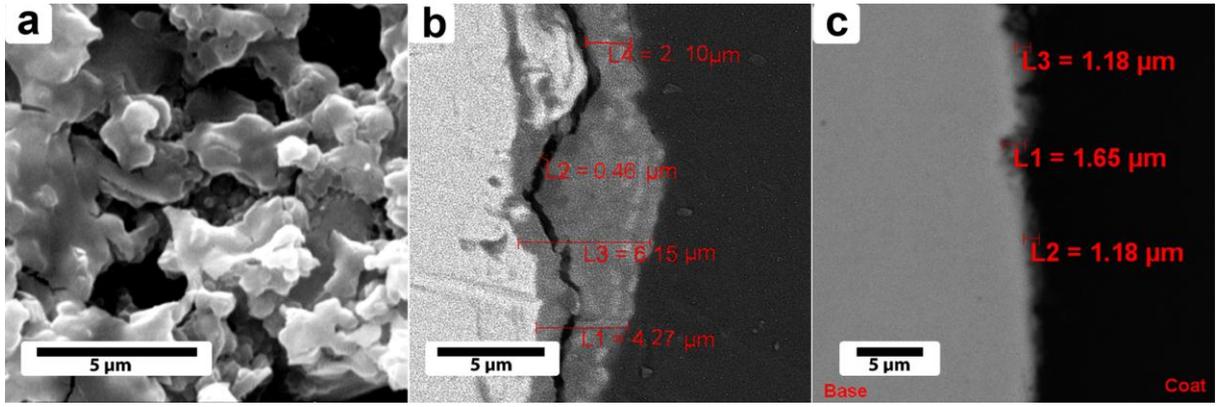

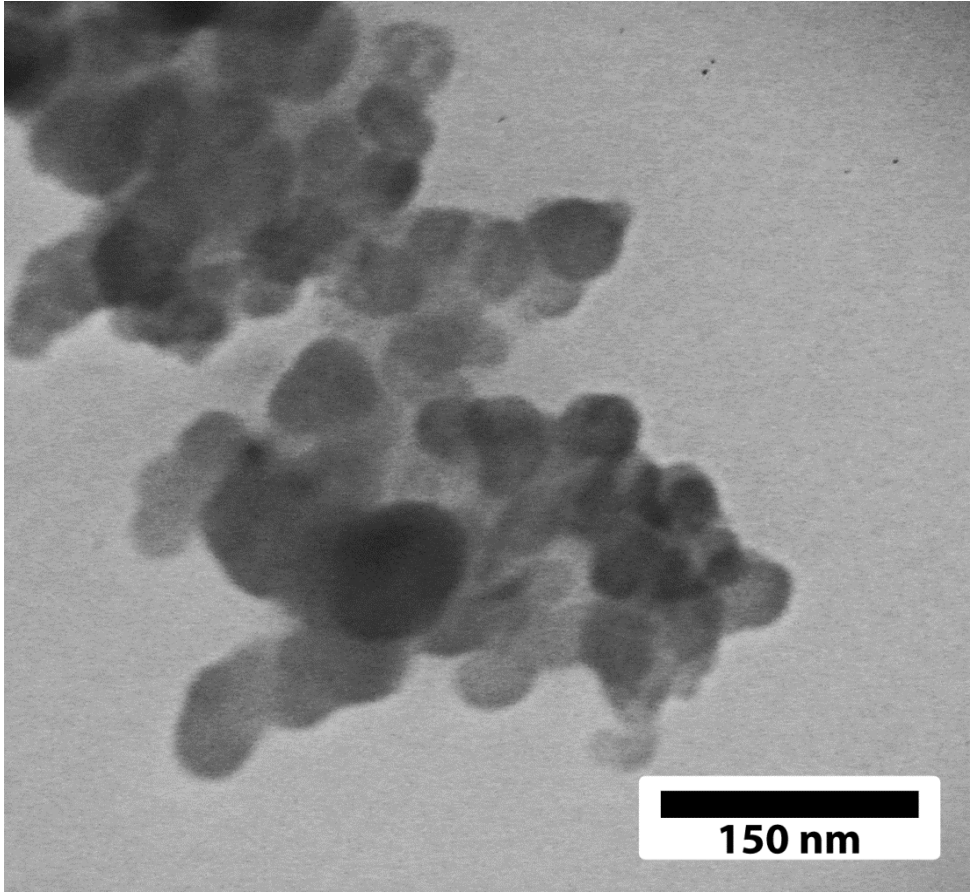

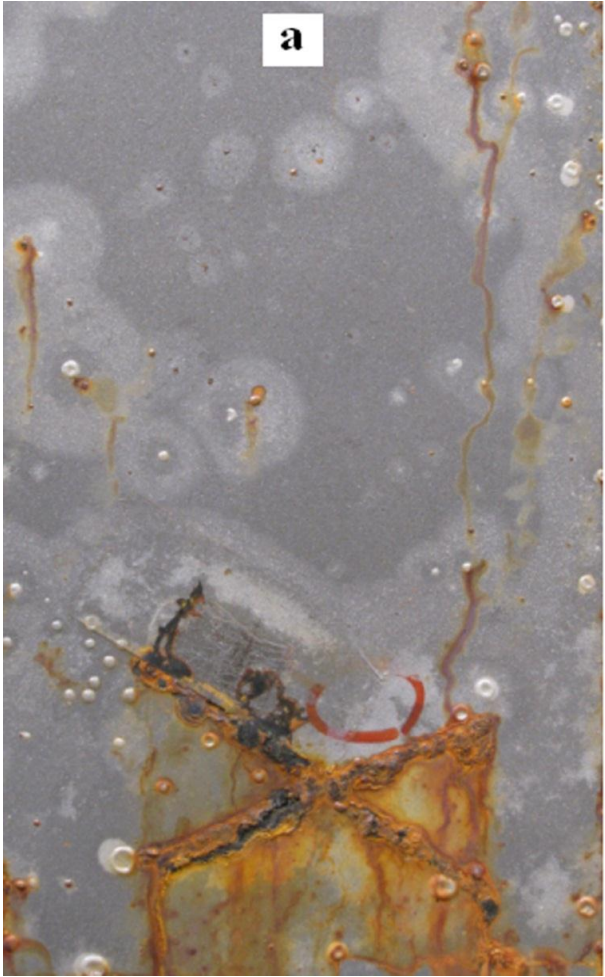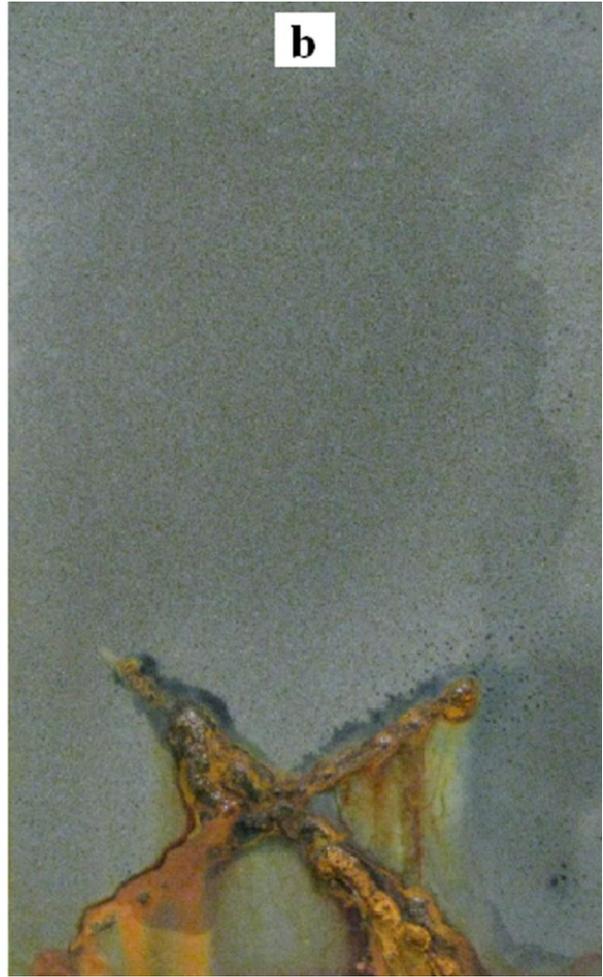

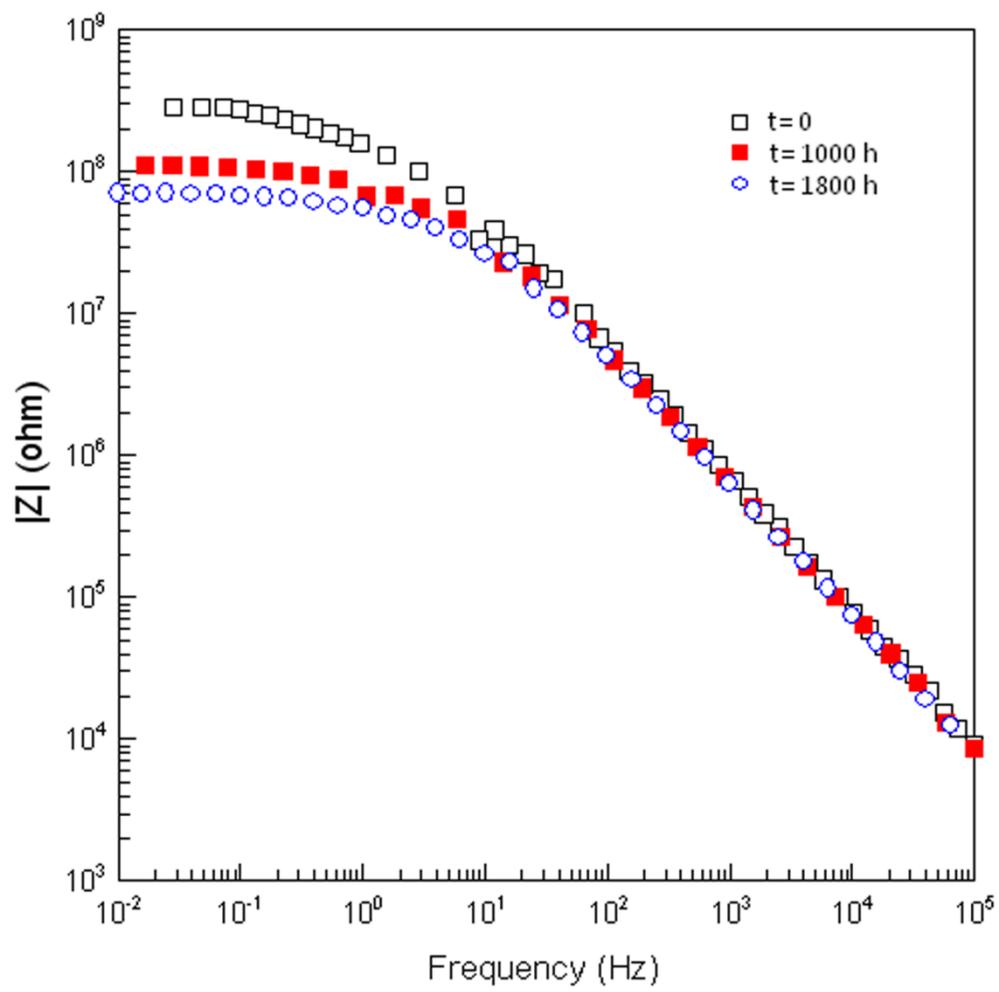

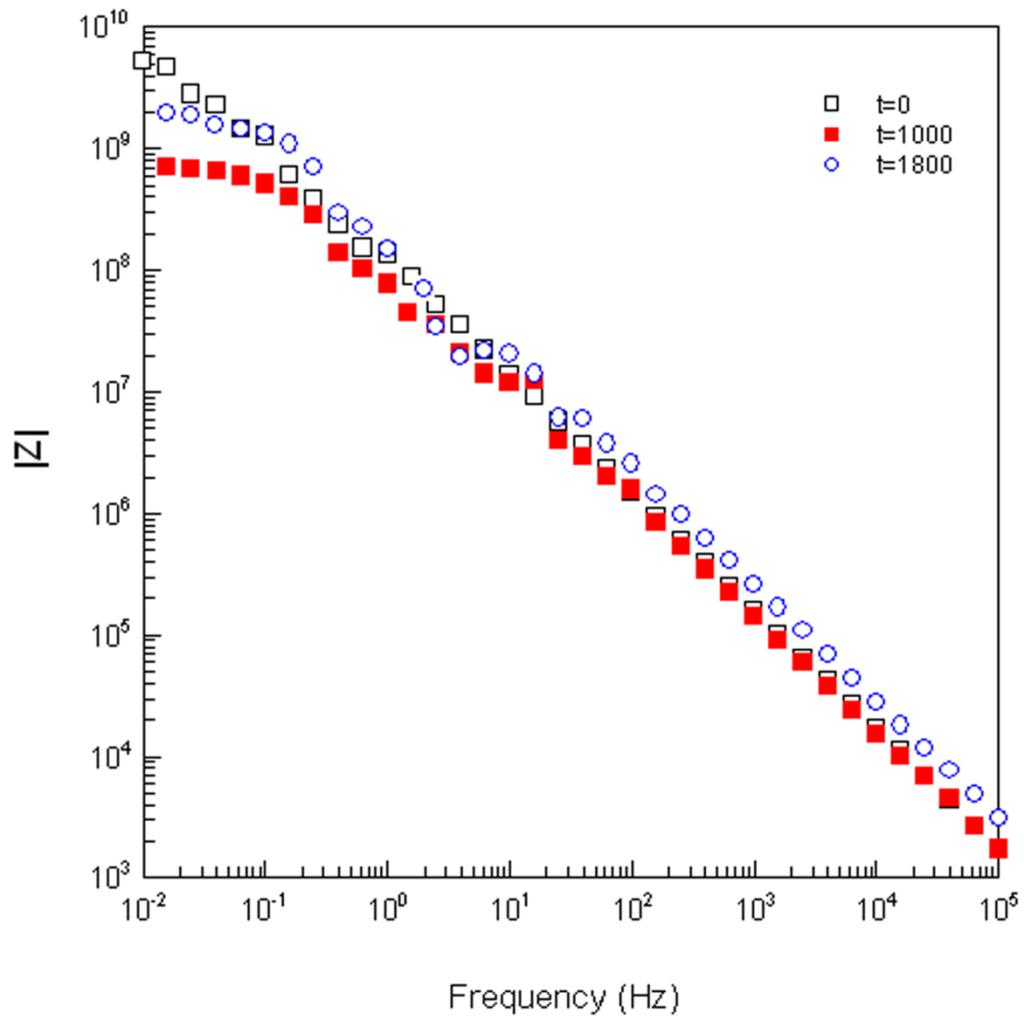

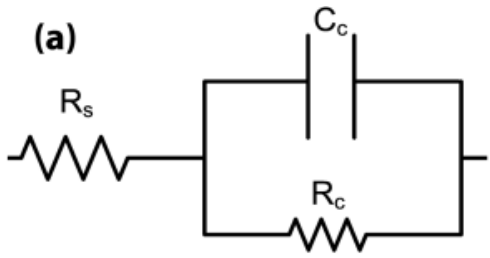 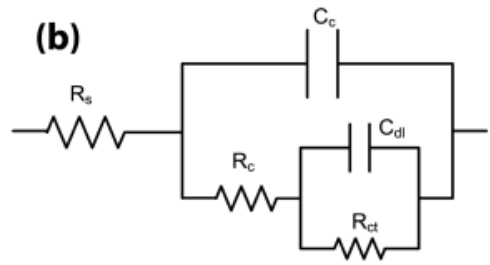

**Supporting Information:**

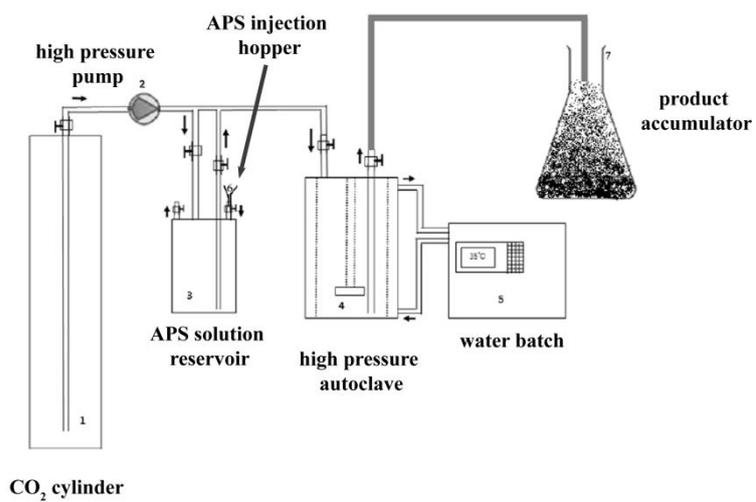

**Figure S1:** Schematic of experimental setup for SPAni production